\documentstyle[preprint,aps,pre]{revtex}

\begin{document}


\title{STOCHASTIC RESONANCE IN SPATIALLY EXTENDED SYSTEMS:
THE ROLE OF FAR FROM EQUILIBRIUM POTENTIALS}

\author{H. S. Wio$^1$ \thanks{Member of CONICET, Argentina;
Electronic Address: wio@cab.cnea.gov.ar}, S. Bouzat$^1$ and and
B. von Haeften$^2$}

\address{1) Grupo de F\'{\i}sica Estad\'{\i}stica
\thanks{http://www.cab.cnea.gov.ar/CAB/invbasica/FisEstad/estadis.htm}
\\ Centro At\'omico Bariloche (CNEA) and Instituto Balseiro (CNEA and
UNC) \\ 8400 San Carlos de Bariloche, Argentina\\ 2) Universidad
Nacional de Mar del Plata, De\'an Funes 3350, 7600 Mar del Plata,
Argentina.}

\date{\today}

\maketitle
\begin{abstract}
Previous works have shown numerically that the response of a
``stochastic resonator'' is enhanced as a consequence of spatial
coupling. Also, similar results have been obtained in a
reaction-diffusion model by studying the phenomenon of stochastic
resonance (SR) in spatially extended systems using {\em
nonequilibrium potential\/} (NEP) techniques. The knowledge of
the NEP for such systems allows us to determine the probability
for the decay of the metastable extended states, and approximate
expressions for the correlation function and the signal-to-noise
ratio (SNR). Here, exploiting known forms of the NEP, we have
investigated the role of NEP's symmetry on SR, the enhancement of
the SNR due to a {\em selectivity\/} of the coupling or diffusion
parameter, and discussed competition between local and nonlocal
(excitatory) coupling.
\end{abstract}



\section{Introduction}

Many recent theoretical and experimental studies have provided a
large body of evidence on the essentially constructive role played
by fluctuations in a variety of intriguing noise induced
phenomena. Some key-examples are: problems of self-organization
and dissipative structures, noise induced transitions, noise
induced {\it phase} transitions, {\it thermal ratchets} or
Brownian motors, coupled Brownian motors, noise sustained
patterns and stochastic resonance  \cite{general1,general2,SR3}.

This last phenomenon, that is {\it stochastic resonance} (SR),
has attracted considerable interest due, among other aspects, to
its potential technological applications for optimizing the
transmission of information in nonlinear dynamical systems. For a
comprehensive recent review see Ref. \cite{SR3}, that shows the
large number of applications in science and technology, ranging
from paleoclimatology, to electronic circuits, lasers, and
noise-induced information flow in sensory neurons in living
systems, to name a few. Several recent papers have aimed at
achieving an enhancement of the output SNR by means of the
coupling of several SR units \cite{buls2,wiocas} in what conforms
an ``extended medium" \cite{wiocas}.

The relevance of pattern-formation phenomena to several areas of
science and technology is a very well established fact. Accounts
of the state of the art can be found in many reviews and books
\cite{general1}, collecting the results obtained over the last
couple of decades regarding the description of pattern formation
and propagation phenomena in self-organizing systems.

It is a common belief that the nontrivial spatio-temporal
behaviour occurring for instance in the Complex Ginzburg-Landau
Equation (CGLE), reaction-diffusion (RD) schemes, and other
nonequilibrium systems, originates from the {\it non-potential}
(or {\it non-variational}) character of the dynamics, meaning
that there is no Lyapunov functional (LF) for the dynamics.
However, Graham and co-workers, have shown that a Lyapunov-like
functional does exist for the CGLE \cite{GR1}.

In order to fix ideas, we start by discussing the existance of LF
in two different dynamical situations. The simplest case, in
which a LF exists, corresponds to a {\it gradient flow system}
such as
\begin{equation}
\label{varia2} \dot x_{j}~=~-\frac{\partial}{\partial x_{j}}
V(x_{1},..,x_{n}).
\end{equation}
The fixed points will correspond to the extrema of the
``potential function" $V(x_{1},..,x_{n})$, and the system will
evolve towards these minima of $V(x_{1},..,x_{n})$ following
trajectories corresponding to the line of {\it steepest descent}.
Clearly $V(x_{1},..,x_{n})$ is a LF as it also satisfies
\begin{equation}
\label{varia3} \frac{d V}{d t}~=~\sum \frac{\partial V}{\partial
x_{j}} \frac{d x_j}{d t}~=~ - \sum \left( \frac{\partial
V}{\partial x_{j}} \right) ^2~\leq~0.
\end{equation}

Now we consider an example corresponding to a non-gradient flow.
We take the following case
\begin{equation}
\label{varia5} \dot x_{j}~=~-\sum ({\cal T })_{jl}\frac{\partial
V}{\partial x_l},
\end{equation}
where ${\cal T }$ is an arbitrary, positive definite matrix. We
can separate it into a symmetric (${\cal S }$) and an
antisymmetric (${\cal A }$) part
\begin{eqnarray}
\label{varia6}
{\cal T }~=~{\cal S }~+~{\cal A } & & \nonumber \\
{\cal S }~=~\frac{1}{2}({\cal T }~+~{\cal T }^T),\,\,\, & &{\cal S
}~= {\cal S }^T \nonumber
\\ {\cal A }~=~\frac{1}{2}({\cal T }~-~{\cal T }^T),\,\,\, & &{\cal A }
~=~-~{\cal A }^T.
\end{eqnarray}
The fixed points are given by the extrema of $V$. On the other
hand we have that $V$ also fulfills
\begin{equation}
\label{varia7} \frac{d V}{d t}~=~\sum ({\cal S
})_{jl}\frac{\partial V}{\partial x_{j}}\frac{\partial
V}{\partial x_{l}} ~-~\sum ({\cal A })_{jl}\frac{\partial
V}{\partial x_{j}} \frac{\partial V}{\partial x_{l}}~\leq~0,
\end{equation}
as, clearly, the first term on the rhs is $\leq 0$, while the
second one is $=0$, hence, $V$ is a LF. The system evolves to the
minima following trajectories different from the steepest descent
ones, determined by ${\cal S }$, since the antisymmetric part of
${\cal T }$ induces a flow in the system that keeps the LF
constant. A thorough (and didactic) discussion on the LF and
different kinds of dynamical flows can be found in \cite{Extend2}.

The use of the LF concept (as well as the related one of
nonequilibrium potential for stochastic systems) in relation to
far from equilibrium systems is scarce. However, there are recent
papers that have made use of these ideas in relation not only
with CGLE \cite{maxi1}, but also for reaction-diffusion systems.
Some of these works concerned the derivation of effective
equations for the evolution of fronts \cite{pego}; while others
were related to the determination of the global stability of the
resulting patterns and the possibility of exchanging the relative
stability between attractors \cite{Extend1,I}.

In the next Section we introduce the concept of nonequilibrium
potential (NEP). When the NEP can be obtained, such an approach
offers an alternative way of confronting a problem that has
attracted considerable attention, both experimentally and
theoretically. Namely, the relative stability of the different
attractors, corresponding to spatially extended states, and the
possibility of transitions among them due to the effect of
fluctuations \cite{maxi1}. The latter aspect which is of great
relevance in the study of SR in spatially extended systems, is
the objective of this work.

The organization of the paper is as follows. As indicated, in the
next section we briefly discuss some basic notions about
nonequilibrium potentials, and show a few relevant examples for
reaction-diffusion systems. In Section III we present the results
for the SNR in some of the previous examples. Section IV contains
a final discussion and some conclusions.

\section{Nonequilibrium Potential}

\subsection{Brief Review}

Loosely speaking, the notion of non-equilibrium potential (NEP)
corresponds to an extension of the notion of equilibrium
thermodynamical potential to non-equilibrium situations. In order
to introduce such NEP, we consider a general form of non-linear
stochastic equations, admitting the possibility of {\em
multiplicative noises}. In particular, we consider equations of
the form
\begin{equation}
\label{EQ} \dot{q}^{\nu}=K^{\nu}(q)+g^{\nu}_{i}(q)\,\xi_i(t),
\qquad\nu=1,...,n;
\end{equation}
where repeated indices are summed over. Equation (\ref{EQ}) is
stated in the sense of It\^{o}. The $\{\xi_i(t),\,(i=1,...,m \leq
n)\}$ are mutually independent sources of Gaussian white noise
with typical strength $\eta$.  The Fokker-Planck equation
corresponding to Eq.(\ref{EQ}) takes the form
\begin{equation}
\label{FP} \frac{\partial P}{\partial,t} =
-\frac{\partial}{\partial q^{\nu}}\,
    K^{\nu}(q)\,P+\frac{\eta}{2}\frac{\partial^2}{\partial q^{\nu}\,
    \partial q^{\mu}}\,Q^{\nu \mu}(q)\,P
\end{equation}
where $P(q,t;\eta)$ is the probability density of observing
$q=(q_1,...,q_n)$ at time $t$ for noise intensity $\eta$, and
$Q^{\nu\mu}(q)=g^{\nu}_{i}(q)\,g^{\mu}_{i}(q)$ is the matrix of
transport coefficients of the system, which is symmetric and
non-negative.  In the long time limit ($t\to\infty$), the
solution of Eq.(\ref{FP}) tends to the stationary distribution
$P_{stat}(q)$. According to \cite{GR1}, $\Phi(q)$, the NEP
associated to Eq.(\ref{FP}), is defined by
\begin{equation}\label{POT}
    \Phi(q)=-\lim_{\eta \rightarrow 0}\eta\,\ln P_{stat}(q,\eta).
\end{equation}
In other words
\begin{displaymath}
    P_{stat}(q)\,d^nq=Z(q)\exp
    \left[-\frac{\Phi(q)}{\eta}+{\cal O}(\eta)\right]\,d\Omega_q,
\end{displaymath}
where $\Phi(q)$ is the NEP of the system and $Z(q)$ is defined as
the limit
\begin{displaymath}
    \ln Z(q)=\lim_{\eta\rightarrow 0}\left[\ln P_{stat}(q,\eta)
    +\frac{1}{\eta}\,\Phi(q)\right].
\end{displaymath}
Here $d\Omega_q=\frac{d^nq}{\sqrt{G(q)}}$ is the invariant volume
element in the $q$-space and $G(q)$ is the determinant of the
contravariant metric tensor (for the Euclidean metric it is
$G=1$). It was shown \cite{GR1} that $\Phi(q)$ is the solution of
a Hamilton-Jacobi like equation (HJE)
\begin{equation}
\label{HaJA}
    K^{\nu}(q)\frac{\partial\Phi}{\partial q^{\nu}}
    +\frac{1}{2}Q^{\nu\mu}(q)\frac{\partial\Phi}{\partial q^{\nu}}
    \frac{\partial\Phi}{\partial q^{\mu}}=0,
\end{equation}
and $Z(q)$ is the solution of a linear first-order partial
differential equation depending on $\Phi(q)$ (not shown here).

Equation (\ref{POT}) and the normalizability condition ensure
that $\Phi$ is bounded from below.  Furthermore, from the
separation of the streaming velocity of the probability flow in
the steady state into conservative and dissipative parts,  it
follows that
\begin{displaymath}
    \frac{d\Phi(q)}{dt}=K^{\nu}(q)\frac{\partial\Phi(q)}{\partial
    q^{\nu}}=-\frac{1}{2}\,Q^{\nu \mu}(q)\,\frac{\partial\Phi}
    {\partial q^{\nu}}\frac{\partial\Phi}{\partial q^{\mu}}\leq 0,
\end{displaymath}
i.e.\ $\Phi$ is a LF for the dynamics of the system when
fluctuations are neglected. Under the deterministic dynamics:
${\dot q}^{\nu}= K^{\nu}(q)$, $\Phi$ decreases monotonically and
takes a minimum value on attractors. In particular, $\Phi$ must be
constant on all extended attractors (such as limit cycles or
strange attractors) \cite{GR1}.

\subsection{Examples of Nonequilibrium Potentials}

\subsubsection{Scalar Reaction-Diffusion Model}

As a first example we focus on a one--dimensional, one--component
model of an electrothermal instability \cite{general1}, which
corresponds to an approximation to the continuous limit of the
coupled system studied by Lindner {\it et al} \cite{buls2}. For
this model, the effect of boundary conditions (b.c.) in pattern
selection, the {\it global stability} of nonhomogeneous
structures, and the critical-like behaviour due to the
coalescence of two patterns \cite{Extend1}, have been studied.

The RD model that we work with describes the time evolution of a
field $\phi (x,t)$ which represents the temperature profile in
the so-called ``hot spot model'' in superconducting microbridges
(or ballast resistor) \cite{Extend1,I}. The evolution of $\phi $
is given by
\begin{equation}
\label{Ballast} \frac{\partial \phi }{\partial t} = D \,
\frac{\partial ^2\phi}{\partial x^2} -\phi + \theta [\phi -
\phi_c] + \xi(x,t),
\end{equation}
where $\xi(x,t)$ is a white noise in space and time, in the
bounded domain $x\in [-L,L]$ with Dirichlet b.c. at both ends,
{\it i.e.} $\phi (\pm L,t)=0$. $\theta [x]$ is the step function.
We restrict our discussion to the parameter range where the
(associated deterministic) system has two stable attractors
(patterns). The piecewise linear approximation of the reaction
term, mimicking a cubic polynomial, was chosen in order to find
analytical expressions for the spatially symmetric solutions of
Eq.(\ref{Ballast}). It is clear that the trivial solution $\phi
_0(x)=0$, which is linearly stable, exists for the whole range of
parameters. Besides this solution we find only one stable
nonhomogeneous structure, $\phi_s(x)$, which presents an excited
($\phi _s(x)>\phi _c$) central zone, and another similar unstable
structure, $\phi _u(x)$, with a smaller excited central zone. The
latter pattern corresponds to the saddle separating both
attractors $\phi _0(x)$ and $\phi _s(x)$ \cite{Extend1,I}.

The indicated patterns are minima of the NEP of our system. For
the present case, such a NEP reads \cite{I}
\begin{equation}
\Phi [\phi ,\phi _c]=\int_{-L}^{+L}\left\{ -\int_0^\phi \left(
-\phi +\theta [\phi -\phi _c]\right) \,d\phi .+\frac D2\left(
\frac{\partial \phi }{\partial x}\right) ^2\,\right\} dx.
\end{equation}
It can be shown that $\frac{\partial \phi }{\partial
t}=-\frac{\delta \Phi }{\delta \phi }$ and also $\dot{\Phi }=-\int
\left( \frac{\delta \Phi }{\delta \phi }\right) ^2dx\leq 0.$ This
functional offers us the possibility to study both the local and
global stability of the patterns as well as the changes
associated to variations of model parameters. \cite{Extend1,I}

In Fig. 1 we depict  $\Phi [\phi ,\phi _c]$ evaluated at the
stationary patterns $\phi _0$ ($\Phi [\phi _0]=0$), $\phi _s(x)$
($\Phi ^s=\Phi [\phi _s]$) and $\phi _u(x)$ ($\Phi ^u=\Phi [\phi
_u]$), for a system size $L=1$, as a function of $\phi _c$ for a
given value of $D$. The upper branch is the NEP for $\phi _u(x)$,
where $ \Phi $ attains an extremum (as a matter of fact it is a
saddle). On the lower branch, for $\phi _s(x)$, and also for
$\phi _0(x)$, the NEP has local minima. The curves exist up to a
certain critical value of $\phi _c$ at which both branches
collapse \cite{Extend1,I}. It is interesting to note that, since
the NEP for $\phi_u(x)$ is always positive and, for $\phi _s(x)$,
$\Phi ^s > 0$ for ``large" values of $\phi _c$ and also $\Phi ^s
< 0 $ for ``small" values of $\phi_c$, $\Phi ^s$ vanishes for an
intermediate value $\phi _c=\phi_c^{*}$, where $\phi _s(x)$ and $
\phi _0(x)$ exchange their relative stability.

\subsubsection{Three Component Activator-Inhibitor Model}

Here we present an exact form of the nonequilibrium potential for
a three component system of the activator-inhibitor type (with
one activator and two inhibitors) in a particular parameter
region. Such a three component system provides the adequate
framework for a minimal model describing pattern formation in
high-pressure or low-pressure chemical reactors \cite{Extend11}.
The system we consider is described by the following set of
equations
\begin{eqnarray}
\label{sys3} \frac{\partial u(x,t)}{\partial t}&=& D \nabla^2
u(x,t) + f(u(x,t)) - v(x,t) - w(x,t)+ g^{u}_1 \xi_1(x,t) +
g^{u}_2 \xi_2(x,t) \nonumber \\
\epsilon_1 \frac{\partial v(x,t)}{\partial t}&=&\nu_1 \nabla^2
v(x,t)+ \beta u(x,t) - \gamma v(x,t)+ g^{v}_1 \xi_1(x,t) +
g^{v}_2 \xi_2(x,t) \nonumber \\ \epsilon_2 \frac{\partial
w(x,t)}{\partial t}&=&\nu_2 \nabla^2 w(x,t) + \beta' u(x,t) -
\gamma' w(x,t) + g^{w}_1 \xi_1(x,t) + g^{w}_2 \xi_2(x,t),
\end{eqnarray}
where $x$ represents an n-dimensional spatial coordinate. The
$\xi_i(x,t)$'s are gaussian white-noise sources of zero mean
satisfying
\begin{equation}
\langle \xi_i(x,t) \xi_j(x',t') \rangle = \eta \delta_{i j}
\delta(t-t') \delta(x-x'),
\end{equation}
where $\eta$ is again a small parameter measuring the noise
intensity. All the parameters and fields shall be considered as
dimensionless (scaled) quantities. We shall only consider
situations where the noise terms in the third of Eqs.\
(\ref{sys3}) are negligible, and we set $g^{w}_1=g^{w}_2=0$.
Furthermore, we will analyze the system in the limit
$\nu_1=\epsilon_2=0$ with $\epsilon_1=1$ and $\nu_2=\nu$, where
for the now {\it temporally slaved inhibitor} $w$ we have
\begin{equation}
\label{wGu} w(x,t)=\int dx' G(x,x') u(x',t),
\end{equation}
where $G(x,x')$ is the Green function of the third of Eqs.
(\ref{sys3}) in the indicated limit \cite{Extend1}. Hence the
system can be reduced to an effective two component system with a
{\it nonlocal interaction} \cite{Extend1,Extend11}.

When the relation
\begin{equation}
\gamma=\frac{Q_u \beta +Q_v}{2 Q_{u v}},  \label{gamma}
\end{equation}
where $Q_u = (g^u_1)^2+(g^u_2)^2, \, Q_v = (g^v_1)^2+(g^v_2)^2$
and $Q_{u v}=g^u_1 g^v_1+g^u_2 g^v_2$, holds, the two equations
for such an effective two component system adopt the form
\begin{equation}
\label{uTFu}
  \left( \begin{array}{c}
                          \frac{\partial u(x,t)}{\partial t} \\
                          \frac{\partial v(x,t)}{\partial t}
           \end{array}
   \right) =
   - {\cal T } \left( \begin{array}{c} \frac{\delta \Phi}{\delta u(x,t)}  \\
                            \frac{\delta \Phi}{\delta v(x,t)}
                        \end{array} \right)
+ \left( \begin{array}{c} g^{u}_1 \xi_1(x,t) + g^{u}_2 \xi_2(x,t)  \\
g^{v}_1 \xi_1(x,t) + g^{v}_2 \xi_2(x,t) \end{array}  \right).
\end{equation}
Here the matrix ${\cal T}$, which has the form of the matrix
${\cal T}$ in Eq. (\ref{varia6}) is given by
\begin{equation}
\label{T}
{\cal T}= \frac{1}{2} \left( \begin{array}{cc} Q_u & 2 Q_{u v} \\
                             0 & Q_v
    \end{array} \right) = \frac{1}{2} \left( \begin{array}{cc} Q_u & Q_{u v} \\
                                                             Q_{u v} &  Q_v
    \end{array} \right) + \left( \begin{array}{cc} 0 & \frac{Q_{u v}}{2} \\
                                               -\frac{Q_{u v}}{2} & 0 \end{array} \right)
= {\cal S} + {\cal A},
\end{equation}
and the functional $\Phi[u(x,t),v(x,t)]$ by
\begin{equation}
\label{Phi}
\Phi[u(x,t),v(x,t)]=\int dx \left[\frac{D}{Q_u}
(\nabla u(x,t))^2 + V(u(x,t),v(x,t)) + \frac{1}{Q_u} \int
dx'G(x,x') u(x,t) u(x',t) \right],
\end{equation}
where
\begin{equation}
V(u,v)=- \frac{2}{Q_u} \int^{u} f(u') du' + \frac{2 Q_{u v}
\beta}{Q_u Q_v} u^2 + \frac{\gamma}{Q_v} v^2 - 2
\frac{\beta}{Q_v} u v.
\end{equation}

When the symmetric matrix $S$ is positive definite (when $Q_u Q_v>
Q_{u v}^2$ holds), the functional $\Phi$, that in the associated
deterministic system decreases monotonously with time, is the NEP
of the system in Eqs. (\ref{uTFu}) satisfying the HJE (Eq.
(\ref{HaJA})). Equation (\ref{gamma}), which resembles a detailed
balance condition, arises in this context as a mathematical
constraint necessary for $\Phi$, as defined by Eq.\ (\ref{Phi}),
to be the solution of the HJE above mentioned.

We limited the analysis to the parameter region where Eq.\
(\ref{gamma}) is valid, the matrix $S$ is positive definite, and
hence the nonequilibrium potential is given by Eq.\ (\ref{Phi}).
Although these conditions impose restrictions on the range of
application of our treatment, it is worth noting that, after
choosing the values of the $g_i^\nu$'s satisfying the condition
of positive definiteness, there is still a wide spectrum of
interesting situations to analyze \cite{Extend11,Extend12}.
Furthermore the nonlinear function $f(u)$ is still arbitrary,
opening a wealth of possibilities. When plotting
$\Phi$vs.$\phi_c$, with $\Phi$ evaluated on the stationary
patterns, we see a result similar to the one shown in Fig. 1.

\section{Stochastic Resonance in Extended Media}

\subsection{Preliminaries }

The calculation of the SNR proceeds, for the spatially extended
problem, through the evaluation of the space-time correlation
function $\langle\phi(y,t)\phi(y^{\prime},t^{\prime})\rangle$. To
do that we have used a simplified point of view, based on the
two-state approach \cite{2state}, which allows us to apply some
known results almost directly. We consider a random system
described by a discrete dynamical variable $x$ adopting two
possible values: $c_1$ and $c_2$, with probabilities $n_{1,2}(t)$
respectively. Such probabilities satisfy the condition
$n_{1}(t)+n_{2}(t)=1$. The equation governing the evolution of
$n_1(t)$ (with a similar one for $n_2(t)=1-n_1(t)$) is
\begin{equation}
\label{master} \frac{dn_{1}}{ dt} = -\frac{d n_{2}}{ dt} =
W_{2}(t)n_{2}(t) - W_{1}(t)n_{1}(t) = W_{2}(t) - [ W_{2}(t) +
W_{1}(t) ] n_{1} ,
\end{equation}
where the $W_{1,2}(t)$ are the transition rates {\it out of} the
$x=c_{1,2}$ states. For the bistable extended system in which we
are interested, such states correspond to the spatially extended
attractors (see Refs.\cite{wiocas,Extend3} for details).

If the system is subject (through one of its parameters) to a
time dependent signal of the form $A \cos (\omega_{s}t) $, up to
first order in the amplitude (assumed to be small) the transition
rates may be expanded as
\begin{eqnarray}
\label{Wdef}
W_{1}(t) &=& \mu_{1}- \alpha_{1} A \cos (\omega_{s} t)   \nonumber \\
W_{2}(t) &=& \mu_{2}+ \alpha_{2} A \cos (\omega_{s} t) ,
\end{eqnarray}
where the constants $\mu_{1,2}$ and $\alpha_{1,2}$ depend on the
detailed structure of the system under study. Here we remark that
the $\mu_i$'s, which are the (time independent) values of the
$W_i$'s without signal, are in general different from each other
as a consequence of the different stability of the two states, and
the same happens to the $\alpha_i$'s. With the indicated
modulation the system becomes nonstationary but we make an
adiabatic assumption considering small signal frequencies that
makes the NEP valid at each time for the corresponding value of
the signal.

Using Eq.\ (\ref{Wdef}) we can integrate Eq.\ (\ref{master}) with
the initial condition $x(t_{0})=x_{0}$ and obtain the conditional
probability $n_1(t\mid x_{0},t_{0})$. This result allows us to
calculate the autocorrelation function, the power spectrum and
finally the SNR. The details of the calculation were shown in Ref.
\cite{wiocas,Extend3}. When the symmetrical case is considered
all the results reduce to those in \cite{2state}. For the SNR, up
to the relevant (second) order in the signal amplitude $A$, we
find for the SNR the result \cite{Extend3}
\begin{equation}
\label{RTilde} {\cal R}=\frac{A^2 \pi (\alpha_2 \mu_1 +\alpha_1
\mu_2)^2}{4\mu_1\mu_2(\mu_1+\mu_2)}.
\end{equation}

In order to evaluate the transition rates between both states we
discretize the space and the fields as
\begin{equation}
x \rightarrow x_i,  \,\,\,\,\,\,  (u(x),v(x)) \rightarrow
(\tilde{u}_1,\tilde{u}_2...,\tilde{u}_N,
\tilde{v}_1,...,\tilde{v}_N)
\end{equation}
and use the Kramers-like formula \cite{Hanggi}
\begin{equation}
\label{w=pref*exp} W_{U_i \rightarrow U_j}\equiv
W_{i}=\frac{\lambda_{+}}{2 \pi}
\sqrt{\frac{\Phi''_i}{|\Phi''_m|}}
\exp{\left[-\frac{(\Phi_m-\Phi_i)}{\eta}\right]},
\end{equation}
where $\lambda_{+}$ is the unstable eigenvalue of the
deterministic flux at the unstable state $U_m$, $\Phi''_i$ and
$\Phi''_m$ indicate the determinants of the matrix of second
order derivatives of the NEP with respect to the discretized
fields in the states $U_i$ and $U_m$ respectively, and $\Phi_i$
and $\Phi_m$ are the values of the NEP evaluated at the stationary
states $U_i$ and $U_m$, $i=1,2$. Finally, in order to compute the
SNR as indicated above, we calculate the parameters $\mu_i$ and
$\alpha_i$ numerically  as
\begin{equation}
\mu_{i}=W_i|_{S(t)=0} \,\,\,\,;\,\,\,\,\, \alpha_{i}=-\left
.\frac{d W_i}{d S(t)}\right |_{S(t)=0}.
\end{equation}

\subsection{Role of the NEP symmetry}

To fix ideas we consider the system given by Eq.\ (\ref{uTFu}) in
one dimension, with the spatial coordinate $x$ varying between
$-L$ and $L$, and assuming Dirichlet boundary conditions for the
three fields. We adopt the following piecewise linealized form
for the nonlinear function $f(u)$
\begin{equation}
f(u)=-u+ \theta (u-a)
\end{equation}
where $\theta (u)$ is the step function. We fix $g^u_1=0,
g^u_2=.02, g^v_1=.01$ and $g^v_2=1$. This leads us to a situation
in which we have essentially only one noise source ($\xi_2$)
acting on $v$. With this choice we have $Q_u \ll Q_v$ and
$\gamma$ results approximately independent of $\beta$.

The piecewise linearized form of $f(u)$ allows us to calculate
analytically the stationary inhomogeneous patterns  of the
associated deterministic system as linear combinations of
exponentials \cite{Extend1,I}. We search for solutions symmetric
with respect to $x=0$, with a central activated region ($u>a$)
surrounded by a non activated region ($u<a$). Depending on the
values of $a, D, \beta$ and $\beta'$ we find four (two stable and
two unstable), two (one stable and one unstable) or zero
stationary inhomogeneous solutions. Furthermore the homogeneous
null solution ($u=v=w=0$) always exists and is stable. We call
these solutions $s_1$ and $ns_1$ (existing in the regions were
there are two or four inhomogeneous solutions), $s_2$ and $ns_2$
(existing only in the region of four solutions) and $s_0$ (the
null solution). The $s_i$'s are the stable solutions and the
$ns_i$'s are the unstable ones.

We will focus our analysis on a region of parameters where the
system has two stationary stable patterns (stationary lineary
stable solutions of Eqs. (\ref{sys3}) for $u,v,$ and $w$) and one
stationary unstable pattern (stationary lineary unstable solution
of Eqs. (\ref{sys3})). In Fig. 3 of Ref.\cite{Extend11} the
$u$-fields for the three patterns for a particular choice of the
parameters were shown. We call $U_1$ the large stable pattern
which has a central activated region ($u>a$), $U_2$ the small
stable pattern which reduces to the homogeneous null solution
when $S$ is set equal to zero, and $U_m$ the unstable pattern. A
complete study of the pattern formation of this system can be
found in Ref. \cite{Extend11}.

In the region of only two stable patterns we are considering, the
deterministic dynamics given by Eqs. (\ref{uTFu}) drives the
system toward one of the patterns (selected depending on the
initial condition) which is reached asymptotically. If small
fluctuations are present in the system the fields fluctuate
around one of the stable patterns and transitions between the two
patterns become possible. Note that the $g^{\mu}_i$ in Eqs.
(\ref{uTFu}) are constants that couple the noises to the system
while the intensity of the fluctuations is determined by the
parameter $\eta$.

It is worth mentioning that the NEP given in Eq. (\ref{Phi}) is
valid for the system in Eq. (\ref{uTFu}) for arbitrary number of
spatial dimensions, for an arbitrary nonlinear function $f(u)$,
and with the parameter region of validity being independent of
the choice of $f(u)$ \cite{Extend11}. The consideration of only
one spatial dimension and the particular election of $f(u)$ are
in order to simplify the calculations, particularly regarding
pattern formation. The signal is introduced as a (slow)
modulation in a parameter $S$ by setting $S=S(t)=A \,
\cos(\omega_s t)$ added to the autocatalytic function $f(u)$.

We now analyze the SR phenomenon in our spatially extended system
using the theory discussed before. To proceed with such an
analysis we identify the two stable patterns ($U_1$ and $U_2$)
with the states of the two-state-theory. Hence, the discrete
variable $x$ will adopt values $c_1$ and $c_2$ according to
whether the system is in the states $U_1$ or $U_2$, yielding the
result for the SNR in Eq. (\ref{RTilde}). The changes induced in
the patterns and their stability by the variation of some model
parameter will be reflected in changes in the values of $\mu _i$
and $\alpha _i$ and, accordingly, will affect the results for the
SNR.

We fix $L=1, \beta=\beta'=1, \gamma=10.026, \gamma'=\nu=10,
g^u_1=1, g^u_2=0, g^v_1=.05$ and $g^v_2=.01$, and leave $D$, $a$
and $\eta$ (the noise intensity) as free parameters. Note that
with the chosen values for the $g^{\mu}_i$'s, the only relevant
noise term in the system (Eq.\ (\ref{uTFu})) is $g^u_1 \xi_1(t)$
in the equation for $u$ that appears added to the signal (hence
it can be considered as coming together with the signal). The
parameters $g^v_1$ and $g^v_2$ are set different from zero to
keep the system inside the parameter region where $\Phi[u,v]$ as
defined in Eq.\ (\ref{Phi}) is valid as a nonequilibrium
potential \cite{Extend3}.

In Fig. 2 we show the results for the SNR $(R)$ as a function of
the noise intensity for different values of $D$ and $a$. We see
that while keeping $a$ constant ($a=.25$) (Fig. 2 a), the largest
values of $R$ are those for $D=D_s$, which is the symmetric
situation. Also, if we fix $D=D_s$ (Fig. 2 b), any departure of
$a$ from the value $.25$  (that is any departure from the
symmetric situation) lowers the values of $R$. Hence, the
symmetric situation is found to be the most favorable one
concerning the improvement of SNR. Note that the maximum of the
$R$ vs. $\eta$ curve (for fixed values of $a$ and $D$), that we
will call $R_{max}$, increases with symmetry and reaches its
largest value for the symmetric situation. In Fig. 3 we show
$R_{max}$  plotted as a function of $D$ for $a=.25$, where it is
apparent that the optimum value of diffusion is $D=D_s$
corresponding, as indicated, to the symmetric case.

A fact that arises from these results is that, while keeping all
the other parameters of the system fixed, there exists an optimal
value of diffusion (coupling of the distributed system) that
maximizes SNR. The interesting aspect is that this optimal value
is the one that makes the potential symmetric.

It is worth mentioning that these results do not contradict but
complete those in \cite{wiocas} where enhancement due to coupling
was found, since in that work only symmetric situations were
analyzed. Roughly speaking, the main result in \cite{wiocas} can
be summarized saying that, given two different symmetric
situations (each one necessarily having different values of $D$
and $a$), the one with the higher value of $D$ produces higher
values of SNR. However, we must keep in mind that for a too large
value of $D$, some of the approximations involved in the
calculations may break down \cite{wiocas}. Also, it is worth here
pointing out that the same thing may happen for too large
asymmetries. For example, consider an extremely asymmetric
situation where the barrier for, say the transition from state
$U_1$ to $U_2$, is much larger than the barrier for the opposite
transition. In such a case, the values of the noise intensity
leading to reasonable jumping rates from $U_1$ to $U_2$ will be
far beyond the validity of the Kramers-like approximation for the
inverse transition.

\subsection{Enhancement due to selective coupling}

The model we discuss next is an extension of the one-component RD
model discussed in Section 2, but now the diffusive parameter
depends on the field $\phi(x,t)$. As a matter of fact, since in
the ballast resistor the thermal conductivity is a function of
the energy density, the resulting equation for the temperature
field includes a temperature-dependent diffusion coefficient in a
natural way.

By adequate rescaling of the field, space-time variables and
parameters, we get a dimensionless time-evolution equation for the
field $\phi (x,t)$
\begin{equation}
\label{Ballast2}
\partial_t\phi(x,t)=\partial_x\left(D(\phi)\partial_x\phi\right) +
f(\phi) + \xi(x,t),
\end{equation}
where $\xi(x,t)$ is a white noise in space and time, and
$f(\phi)=-\phi+\theta(\phi-\phi_c)$, $\theta(x)$ is the step
function.  All the effects of the parameters that keep the system
away of equilibrium (such as the electric current in the
electrothermal devices or some external reactant concentration in
chemical models) are included in $\phi_c$.

As was done for the reaction term \cite{wiocas,Extend1,I}, a
simple choice that retains however the qualitative features of the
system is to consider the following dependence of the diffusion
term on the field variable
\begin{equation}
\label{diff} D(\phi)~=~D_0(1+h\,\theta [\phi-\phi_c)],
\end{equation}
For simplicity, here we choose the same threshold $\phi _c$ for
the reaction term and the diffusion coefficient.

We assume the system to be limited to a bounded domain $x \in
[-L,L]$ with Dirichlet boundary conditions at both ends, i.e.\
$\phi(\pm L,t)=0$. As before, the piecewise-linear approximation
of the reaction term in Eq.(\ref{Ballast2}) was chosen in order to
find analytical expressions for its stationary
spatially-symmetric solutions.  In addition to the trivial
solution $\phi_0(x)=0$ (which is linearly stable and exists for
the whole range of parameters) we find another linearly stable
nonhomogeneous structure $\phi_s(x)$---presenting an excited
central zone (where $\phi_s(x)>\phi_c$) for $-x_c\le x\le
x_c$---and a similar unstable structure $\phi_u(x)$, which
exhibits a smaller excited central zone. The form of these
patterns is analogous to what has been obtained in previous
related works \cite{Extend1,I}.  The difference is that in the
present case $d \phi/d x|_{x_c}$ is discontinuous and the area of
the ``activated" central zone depends on $h$.

The indicated patterns are extrema of the NEP.  In fact, the
unstable pattern $\phi_u(x)$ is a {\em saddle-point\/} of this
functional, separating the {\em attractors\/} $\phi_0(x)$ and
$\phi_s(x)$.  For the case of a field-dependent diffusion
coefficient $D(\phi(x,t))$ as described by Eq. (\ref{Ballast2}),
the NEP reads \cite{Extend3}
\begin{equation}
\Phi [\phi]=\int_{-L}^{+L}\left\{-\int_0^\phi
D(\phi')f(\phi')\,d\phi'+
\frac{1}{2}\left(D(\phi)\frac{\partial\phi}{\partial
x}\right)^2\,\right\}dx.
\end{equation}
Given that $\partial_t\phi~=~-(1/D(\phi)){\delta \Phi}
/{\delta\phi}$ one finds $\dot{\Phi }=-\int\left({\delta\Phi
}/{\delta\phi}\right)^2dx\leq 0$, thus warranting the LF property.

For a given threshold value $\phi _c^*$, both wells corresponding
in a representation of the NEP to the linearly stable states have
the same depth (i.e.\ both states are equally stable).  Figure 4
shows the dependence of $\Phi [\phi]$ on the parameter $\phi
_c$.  As in previous cases, we analyze only the neighborhood of
$\phi _c=\phi _c^*$ \cite{wiocas}.  Here we also consider the
neighborhood of $h =0$, where the main trends of the effect can
be captured.

In Fig.5 we depict the dependence of $R$ on the noise intensity
$\gamma $, for several (positive) values of $h$.  These curves
show the typical maximum that has become the fingerprint of the
stochastic resonance phenomenon.  Figure 6 is a plot of the value
$R_{max}$ of these maxima as a function of $h$.  The dramatic
increase of $R_{max}$ (several $dB$ for a {\em small} positive
variation of $h$), is apparent and shows the strong effect that
the selective coupling (or field-dependent diffusivity) has on
the response of the system.

It must be noted that the only two approximations made in order to
derive Eq.(\ref{RTilde})\textemdash namely the Kramers-like
expression in Eq.(\ref{w=pref*exp}) and the two-level
approximation used for the evaluation of the correlation function
\cite{wiocas,Extend3}\textemdash break down for large positive
values of $h$ because for increasing selectivity the curves of
$\Phi [\phi]$ vs.\ $\phi_c$ in Fig.\ 4 shift towards the left,
which in turn means that the barrier separating the attractors at
$\phi_c^{*}$ tends to zero.  This effect is basically the same as
the one discussed in Ref.\cite{wiocas} in connection with global
diffusivity $D_0$. It is also worth noting that except for the two
aforementioned approximations, all the previous results (e.g.\
the profiles of the stationary patterns and the corresponding
values of the non-equilibrium potential) are analytically exact.

\section{Conclusions }

In this work we have presented the basic ingredients for studying
SR in  coupled or extended media ({\it stochastic resonant medium}
\cite{wiocas}) by means of nonequilibrium potential techniques.
We have shown some examples of NEP for RD systems and the way we
can exploit them to get information about the system's response
(the SNR) to weak potential modulations.

In previous works we have shown, in agreement with numerical
simulations \cite{buls2} how the coupling enhances the system's
response. Here we have shown two other aspects: (a) the role
played by the symmetry of the NEP on such enhancement and; (b)
the effect of selective coupling.

In the first case the results we have obtained clearly indicate
the central role played by symmetry in improving the SNR. We
studied the behavior of $R_{max}$, that is the maximum of the SNR
vs. $\eta $ curve, as the different model parameters are (not
simultaneously) varied, finding that $R_{max}$ always increases
with the symmetry of the potential. This fact led us to our main
result: the optimal values of the different model parameters (for
instance diffusivity or threshold), as regards the maximization
of $R_{max}$, correspond to those making the potential more
symmetric in each situation. Besides the analysis of the
influence of symmetry on stochastic resonance, it is important to
remark that the mere consideration of asymmetric situations has
its own relevance. This is because such bistable asymmetric
models provide, for example, the appropriate framework for
describing SR in voltage--dependent ion channels, as proposed in
some biological experiments. In those systems, the conducting
state is associated a higher-energy well than the non--conducting
one \cite{nature}.

In the second case, the results of our analysis of a scalar RD
system (a generalization of the continuous limit of the model
analyzed in \cite{buls2,wiocas}) indicate that a ``selective"
coupling (that is a field diffusion dependent constant) could
offer a new enhancing mechanism in coupling systems. This
prediction prompts to devise experiments (for instance, through
electronic setups) as well as numerical simulations taking into
account the indicated selective coupling.

One direction in which the present studies can be extended
corresponds to the analysis of the competition between local and
nonlocal couplings. Consider the same NEP described by Eq.
(\ref{Phi}), but with only one inhibitor, the temporally slaved
$w$. For this system the NEP $\Phi$ has the form
\begin{equation}
\label{Phi2} \Phi[u(x,t),v(x,t)]=\int dx \left[\frac{D}{2}
(\nabla u(x,t))^2 + C \int dx'G(x,x') u(x,t) u(x',t) \right],
\end{equation}
where the coupling constant $C$, that in one of the systems
studied in \cite{wiocas} was positive (that is due to an
inhibition interaction), is now assumed negative (that is an {\it
excitatory interaction}).

The results of preliminary studies for the case of excitatory
coupling, where we used the same kind of approach indicated
above, are shown in Fig. 6. There we have compared the results
for the SNR vs. noise intensity for the FitzHugh-Naguno model
\cite{wiocas} with inhibitory interaction, with those obtained in
the above indicated case. It is apparent, on one hand, the SNR
enhancement due to the excitatory interaction, while on the
other, that small changes in the diffusion constant of the
activator (``local coupling" constant) strongly affect the SNR.
The analysis of the different aspects of this problem will be
the subject of a forthcoming paper \cite{berni}. \\ \\

{\bf Acknowledgments:} The authors want to thank V. Gr\"unfeld for
a revision of the manuscript. HSW thanks to F.Castelpoggi,
R.Deza, G.Drazer, G.Iz\'us, M.Kuperman, D.Zanette, for their
collaboration in early stages of this work; to R. Graham, E.
Tirapegui and R. Toral for useful discussions; and acknowledges
financial support from CONICET (Project PIP--4953/96).

\newpage

\begin{figure}[tbp]
\caption{NEP $\Phi$, evaluated at the stationary patterns, for
Dirichlet b.c., as a function of $\phi _c$, for $L=1$ and $D=1$.
The bottom curve (1) corresponds to $\phi _s(x)$ and the top one
(3) to $\phi _u(x)$. The bistability point $\phi_c^*$, is
indicated.}
\end{figure}

\begin{figure}[tbp]
\caption{ a) SNR as a function of the noise intensity for $a=.25$
and different values of $D$. The solid line corresponds to the
symmetric situation $D=D_s$, the long-dashed line to $D=.35$ and
the short-dashed line to $D=.25$. b) SNR as a function of the
noise intensity for $D=D_s$ and different values of $a$. The
solid line corresponds to the symmetric situation $a=.25$, the
dotted line corresponds to $a=.27$ and the dot-dashed to $a=.23$.}
\end{figure}

\begin{figure}[tbp]
\caption{Maximum of SNR ($R_{max}$) as a function of the
activator diffusion $D$ for $a=.25$. The maximum of $R_{max}$
occurs for the symmetric situation $D=D_s$}
\end{figure}

\begin{figure}[tbp]
\caption{SNR ($R$) as a function of the noise intensity (here
indicated by $\gamma$), for three values of $h$: $h=0.0$ (full
line), $h=-0.25$ (dashed line) and $h=0.25$ (dotted line). We
have fixed $L=1$, $D_0=1$, $\delta\phi_c=0.01$ and $\Omega=0.01$.}
\end{figure}

\begin{figure}[tbp]
\caption{Maximum $R_{max}$ of the SNR curve (Fig.4) as a function
of $h$, for three values of $D_0$: $D_0=0.9$ (dashed line),
$D_0=1.$ (full line) and $D_0=1.1$ (dotted line).  The arrows
{\bf a} and {\bf b} indicate the response gain due to an
homogeneous increase of the coupling and to a selective one
respectively.  The larger gain in the second case is apparent.
The inset shows the dependence of $R_{max}$ on $D_0$ for
$h=-0.25$ (lower line), $h=0$ and $h=0.25$ (upper line).}
\end{figure}

\begin{figure}[tbp]
\caption{SNR for the case of competition between local and
nonlocal interaction as a function of the noise intensity (here
indicated by $\gamma$). All curves have $D_v = 4$ (nonlocal
interaction). The following curves correspond to a nonlocal
excitatory interaction with different values of $D_u$: the upper
broken line $D_u = 1.2$, the continuous line $D_u = 0.8$, the
lower broken line $D_u = 1$. The dotted line corresponds to the
original (inhibitory) FitzHugh-Nagumo case $D_u = 1.2$).}
\end{figure}

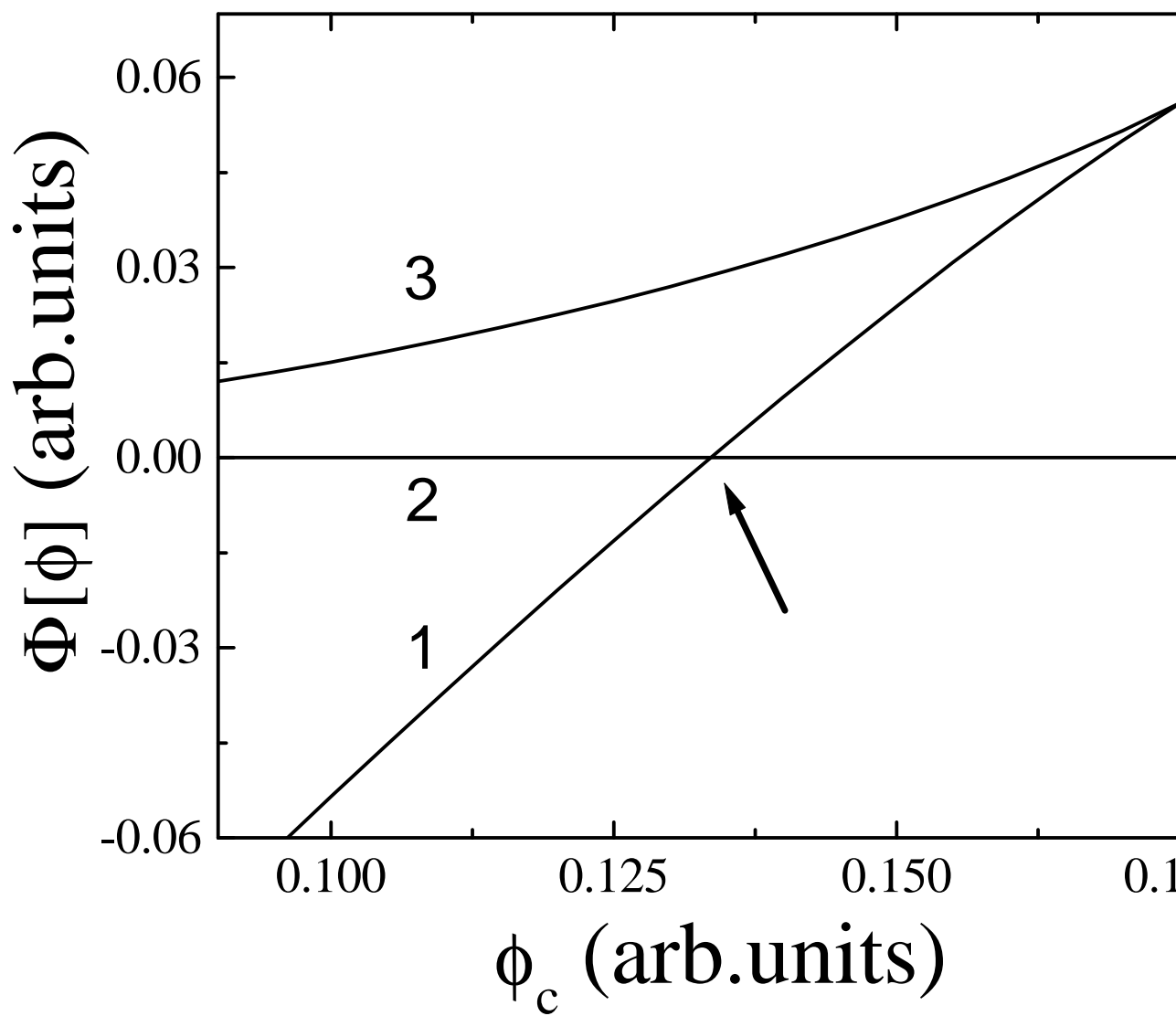
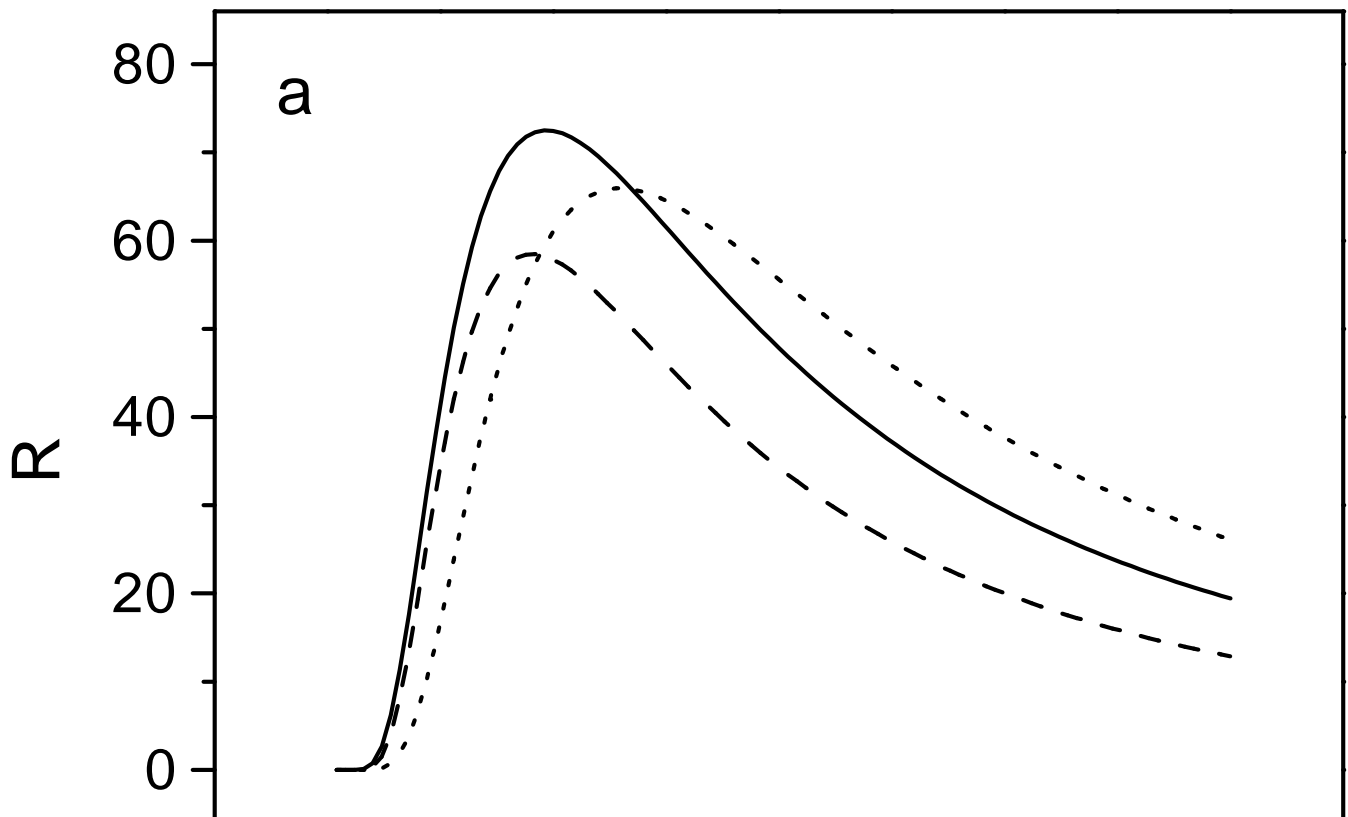
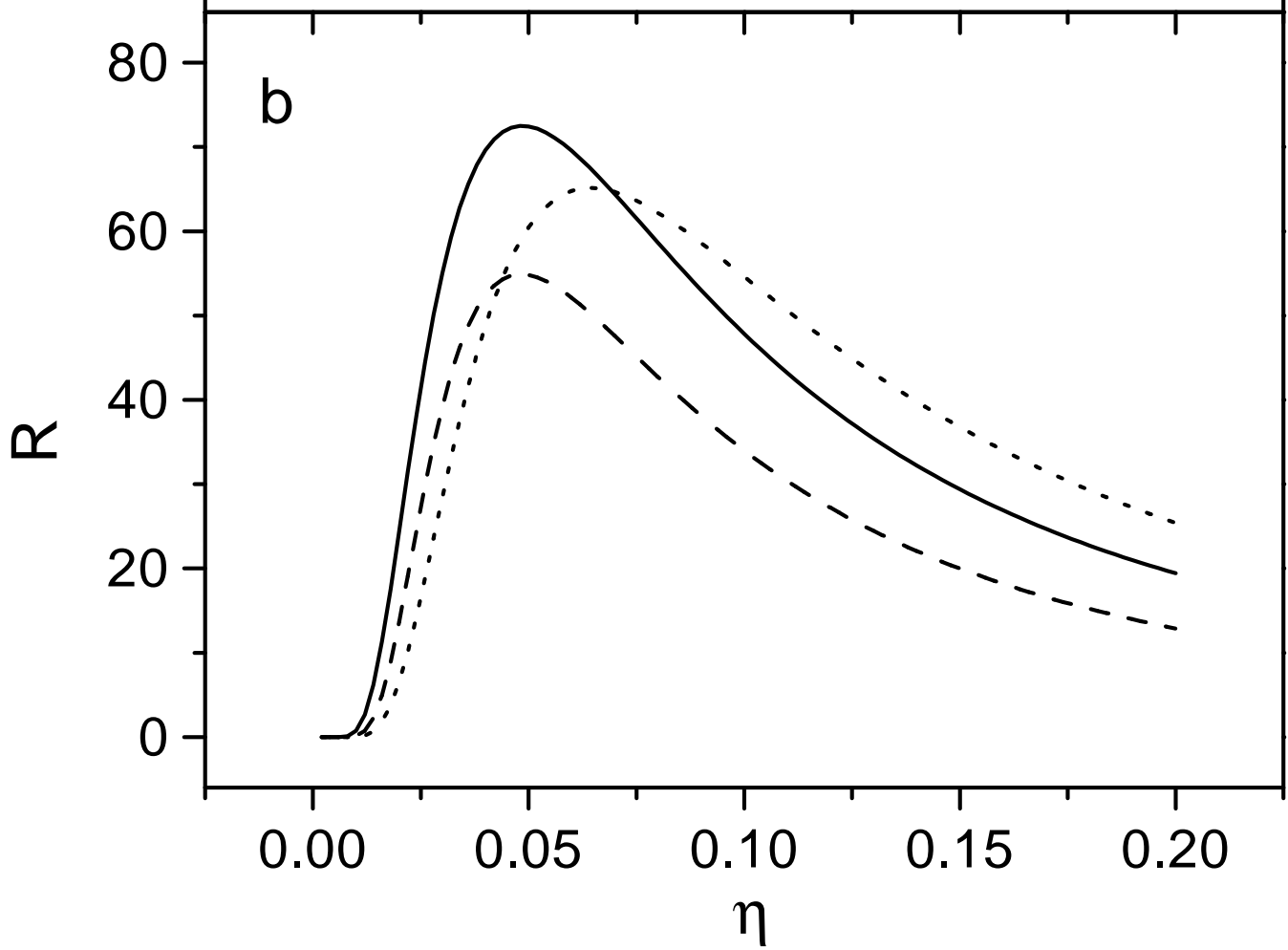
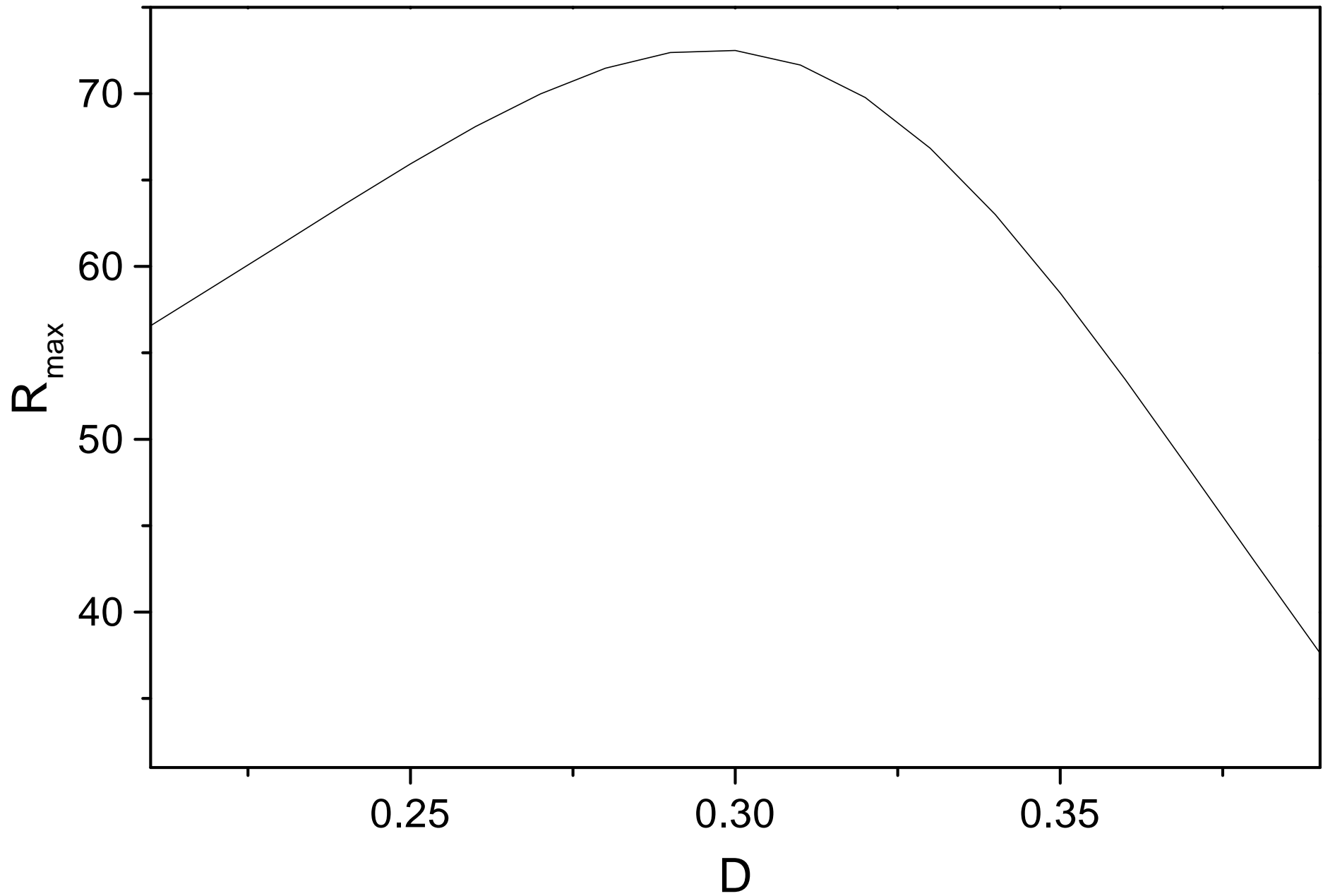
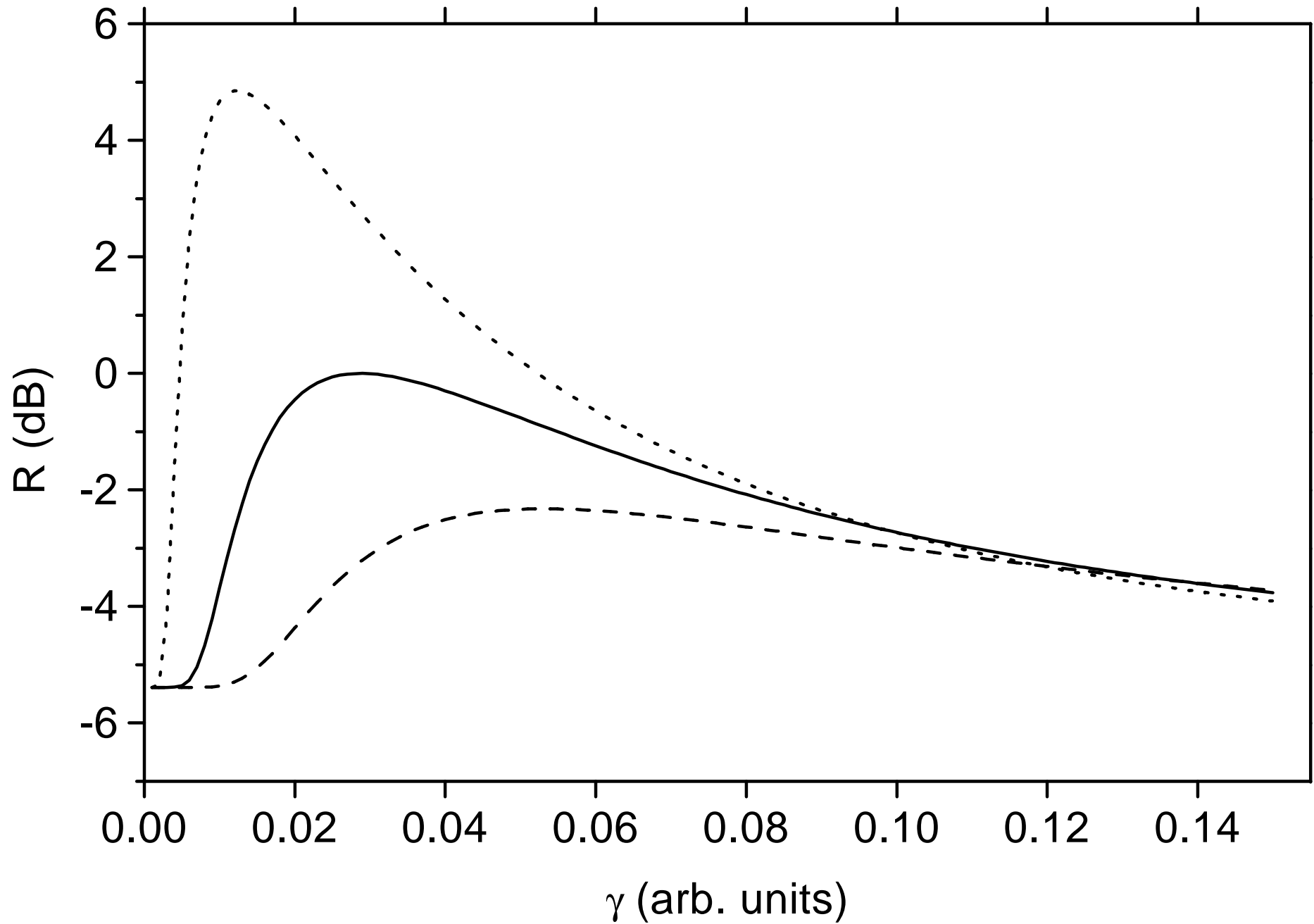
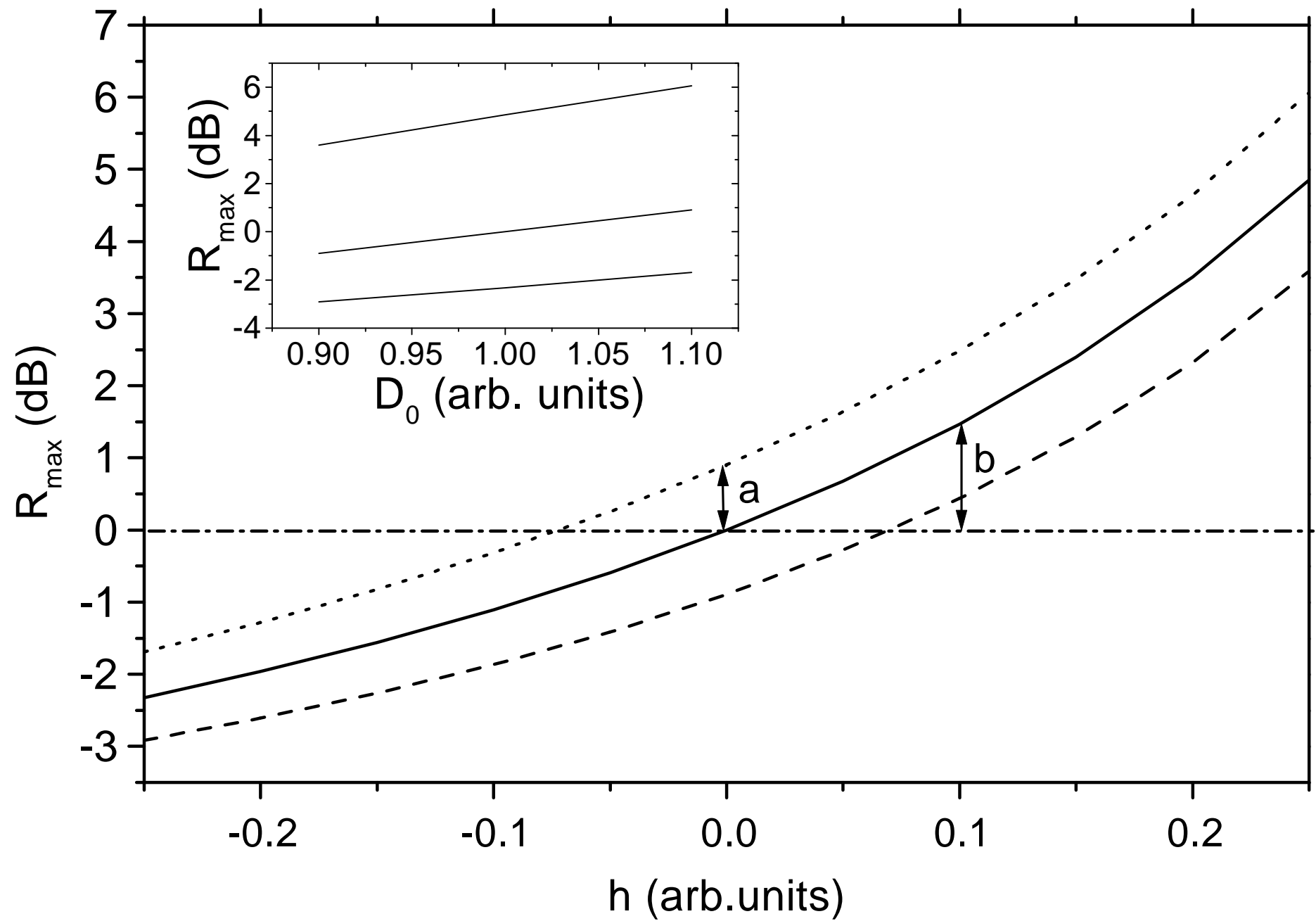
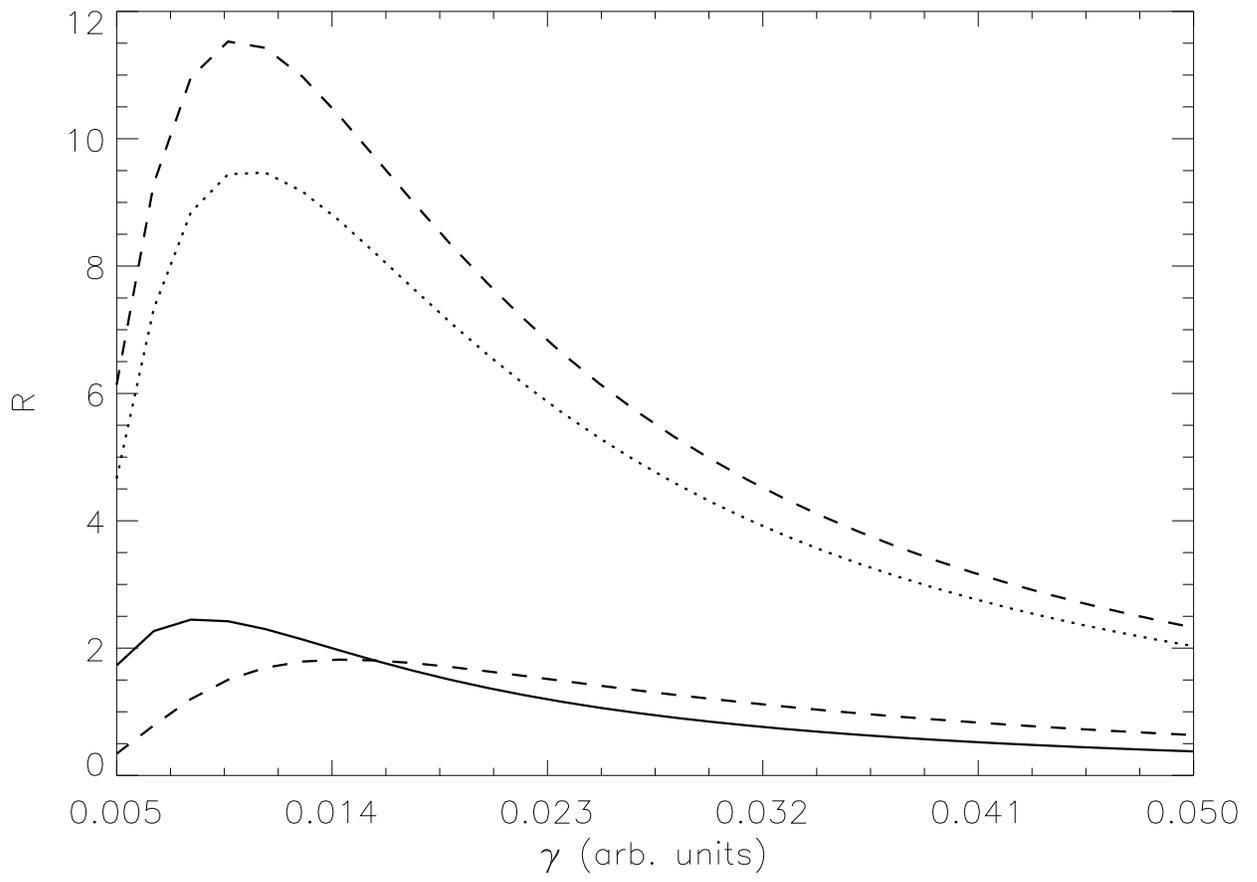
\end{document}